# Software Module Clustering based on the Fuzzy Adaptive Teaching Learning based Optimization Algorithm

Kamal Z. Zamli[1], Fakhrud Din[1,2], Nazirah Ramli[3] and Bestoun S.Ahmed[4]

[1] Faculty of Computer Systems & Software Engineering, Universiti Malaysia Pahang, 26300 Gambang, Kuantan, Pahang, Malaysia

[2] Department of Computer Science & IT, Faculty of Information Technology, University of Malakand, KPK, Pakistan

[3] Department of Mathematics and Statistics, Faculty of Information Technology and Quantitative Science, Universiti Teknologi MARA Pahang, Malaysia

[4] Faculty of Electrical Engineering, Department of Computer Science, Czech Technical University, Czech Republic

**Abstract.** Although showing competitive performances in many real world optimization problems, Teaching Learning based Optimization Algorithm (TLBO) has been criticized for having poor control on exploration and exploitation. Addressing these issues, a new variant of TLBO called Adaptive Fuzzy Teaching Learning based Optimization (ATLBO) has been developed in the literature. This paper describes the adoption of Fuzzy Adaptive Fuzzy Teaching Learning based Optimization (ATLBO) for software module clustering problem. Comparative studies with the original Teaching Learning based Optimization (TLBO) and other Fuzzy TLBO variant demonstrate that ATLBO gives superior performance owing to its adaptive selection of search operators based on the need of the current search.

## 1 Introduction

Software module clustering can be used to assess and facilitate the software comprehension, evolution, and maintenance [1-3]. In fact, evidence shows that modularized software could lead to better development and maintenance process [4, 5]. Recently, researchers have formulated the software module clustering as an optimization problem within the area of Search based Software Engineering.

Many approaches have been adopted to address the software module clustering problem exploiting numerous meta-heuristic algorithms. Hill climbing (HC) has been exploited by Mancoridis and Mitchell for this purpose (i.e. based on a tool called

Bunch [6]). Similarly, HC has also been adopted by Mahdavi et al [7]. Genetic algorithms (GA) have also been used for software module clustering by Kumari and Srivinas [8]. Praditwong et al [9] has pioneered the multi-objective approach for software module clustering by using Pareto optimality concept. More recently, Huang et al [10] has proposed a multi-agent evolutionary algorithm.

Despite some useful progress, the adoption of new developed meta-heuristic algorithm is deemed necessary. As suggested by the *No Free Lunch theorem* [11], no single meta-heuristic can outperform all others even over different instances of the same problem (e.g. [12-17]). For this reason, our work explores the possibility of exploiting the Teaching Learning based Optimization (TLBO) [18] for tackling the software module clustering problem. Specifically, we also compare the effectiveness of TLBO against Fuzzy TLBO [19]. and Adaptive TLBO [20] based on the Mamdani fuzzy implementation.

The rest of the paper is organized as follows. Section 2 describes the software module clustering problem. Section 3 gives an overview of TLBO and its fuzzy adaptive variants. Section 4 presents our benchmark comparison between TLBO, ATLBO and other TLBO variants. Section 5 presents our discussion. Finally, section 6 presents the conclusion of the work.

## 2 Software Module Clustering Problem

Software module clustering problem can be defined as the problem of partitioning modules into clusters based on some predefined quality criterion. Typically, the quality criterion for software module clustering problem relates to the concept of coupling and cohesion. Coupling is a measurement of dependency between module clusters whilst cohesion is the measurement of the internal strength of a module cluster. Thus, a good cluster distribution aids in functionality-cluster-module traceability provides easier navigation between sub-systems and enhances source code comprehension.

To evaluate the cluster distribution, a software system is usually represented as a Module Dependency Graph (MDG) [6]. The coupling of a cluster can be calculated by summing the weight of external edges leaving or entering a cluster partition (termed *inter-edges*). Meanwhile, cohesion is calculated by summing the internal edges where the source and target modules belonging to the cluster partition (termed *intra-edges*). Combining coupling and cohesion, Mancoridis and Mitchell [6] (and later refined by Praditwong et al [9]) define Modularization Quality($MQ$) measure as the sum of the ratio of intra-edges and inter-edges in each cluster, called Modularization Factor ($MF_k$) for cluster $k$. $MF_k$ can be formally defined as follows:

$$MF_k = \begin{cases} 0 & if\ i = 0 \\ \frac{i}{i+\frac{1}{2}j} & if\ i > 0 \end{cases} \qquad (1)$$

where $i$ is the weight of intra-edges and $j$ is that of inter-edges. The term $\frac{1}{2}j$ is to split the penalty of inter-edges across the two clusters that are connected by that edge. The $MQ$ can then be calculated as the sum of $MF_k$ as follows:

$$MQ = \sum_{k=1}^{n} MF_k \qquad (2)$$

when *n* is the number of clusters.

To illustrate the modularization quality (MQ) calculation, Fig. 1 highlights a two cluster modularization of a class diagram (i.e. referred to as MF1 and MF2). In this case, the class relationship is considered two ways when no navigation is specified (i.e. two relationships from source to destination and from destination to source). For the given class diagram in Fig. 1 the most minimum possible cluster is 1 whilst the maximum possible clusters are 6. The goal is to find the clusters from 2 to 5 that maximize the MQ.

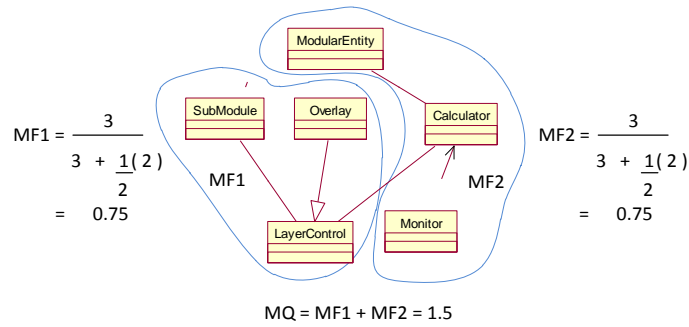

**Fig. 1.** Sample MQ Calculation

## 3 Overview of Adaptive TLBO

Teaching Learning based Optimization (TLBO) [18, 21] takes an analogy from teaching and learning process between teacher and his/her students. Basically, a teacher has more knowledge than the students (i.e. with better fitness value). He or she is trying to impart knowledge to the students so as to match with his/her competency level (refer to the top of Fig. 2). As teachers also have different competency levels, there could also be potential improvements if the students learned from other teachers as well (in any subsequent iterations). At the same time, students can also learn from other students to improve their competency level (refer to the bottom of Fig. 2).

Within TLBO, the solution is represented in the population *X*. An individual $X_i$ within the population represents a single possible solution. Specifically, $X_i$ is a vector with *D* elements where *D* is the dimension of the problem representing the subjects taken by the students or taught by the teacher.

TLBO divides the whole searching process into two main phases; the teacher phase and the learner phase. In order to perform the search, TLBO undergoes the both two phases sequentially one-after-the-other per iteration (see Fig. 3 (a)). The teacher phase involves invoking the global search operation (i.e. exploration). At any instance of the search process, the teacher is always assigned to the best individual $X_i$. The algorithm attempts to improve other individual $X_i$ by moving their position towards $X_{teacher}$ taking into account the current mean value of the population, $X_{mean}$ as follows:

$$X_i^{t+1} = X_i^t + r(X_{teacher} - T_F X_{mean}) \qquad (3)$$

where $X_i^{t+1}$ is the new updated $X_i^t$, $X_{teacher}$ is the best individual in the $X$ population, $X_{mean}$ is the mean of the $X$ population, $r$ is the random number from [0,1] and $T_F$ is a teaching factor which can either be 1 or 2 to emphasize the quality of students.

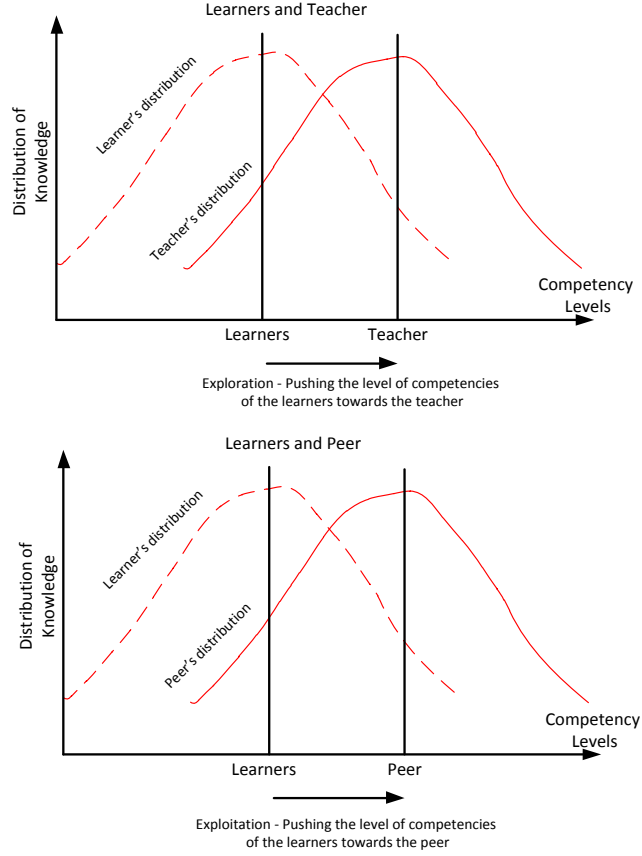

**Fig. 2.** TLBO's Teaching and Learning Analogy [18]

The learner phase exploits the local search operation (i.e. exploitation). Specifically, the learner $X_i^t$ increases its knowledge by interacting with its random peer $X_j^t$ within the $X$ population (i.e. $i \neq j$). A learner learns if and only if the other learner has more knowledge than he or she does. At any iteration $i$, if $X_i^t$ is better than $X_i^t$, then $X_j^t$ is moved toward $X_i^t$ (refer to equation 4). Otherwise, $X_i^t$ is moved toward $X_j^t$ (refer to equation 5)

$$X_i^{t+1} = X_i^t + r\left(X_j^t - X_i^t\right) \qquad (4)$$

$$X_i^{t+1} = X_i^t + r\left(X_i^t - X_j^t\right) \qquad (5)$$

where $X_i^{t+1}$ is the new updated $X_i^t$ , $X_j^t$ is the random peer, and r is the random number from [0,1].

ATLBO [20] is the recent variant of TLBO that adaptively selects its global search (i.e., teacher phase) and local search (i.e., learner phase) operations by using a 3 inputs and 1 output Mamdani-type fuzzy inference system(see Fig. 3 (b)). The inputs are called quality measure $Q_m$, the intensification measure $I_m$ and the diversification measure $D_m$. Equations 6 through 8 correspond to the scaled values of these inputs.

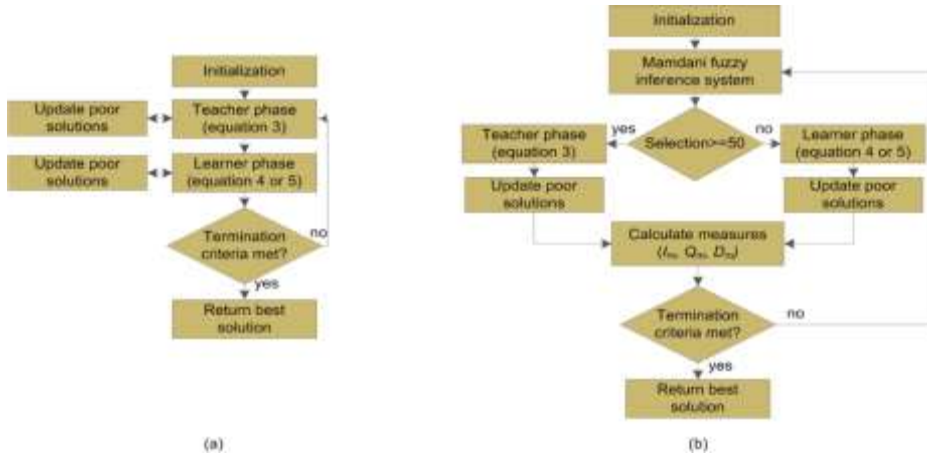

**Fig. 3.** (a) TLBO and (b) ATLBO [20]

$$Q_m = \left[ \frac{X_{current\ fitness} - min\ fitness}{max\ fitness - min\ fitness} \right] \cdot 100 \qquad (6)$$

$$I_m = \left[ \frac{|X_{best} - X_{current}|}{D} \right] \cdot 100 \qquad (7)$$

$$D_m = \left[ \frac{\sum_{j=1}^{population\ size} |X_j - X_{current}|}{D} \right] \cdot 100 \qquad (8)$$

The measure $Q_m$ measures the quality of the current candidate solution. The intensification measure $I_m$ measures whether the current solution is good as compared to the best solution. Finally, the diversification measure $D_m$ measures the current solution against the entire population. The last two measures are based on the Hamming distance. The single output of fuzzy inference system is selection These inputs are transformed into linguistic variables which are fuzzified using trapezoidal membership functions. The rule base of the system is composed of four fuzzy linguistic rules with max-min fuzzy inference method. Finally, center of gravity (COG) is used as defuzzification method to obtain the single crisp output, the selection. This value is then used to decide whether to launch global search or local search rather than executing both as in original TLBO. Fig. 4 summarizes the main component of fuzzy inference system of ATLBO.

## 4 Benchmarking Case Studies

Our benchmarking experiments focus on characterizing the performances of ATLBO against the original TLBO and that of FATLBO [19]. We note that direct comparative performance of ATLBO with the original TLBO (i.e. even with the same number of iterations) and FATLBO can also be unfair. With the same number of iteration, the original TLBO has twice as much fitness function evaluation as compared to ATLBO owing to the serial execution of both exploration and exploitation steps. Furthermore, the number of iterations for both teacher and student phase are also non-deterministic form FATLBO. Hence, for the fair comparison, we opt to use the same number of fitness function evaluations.

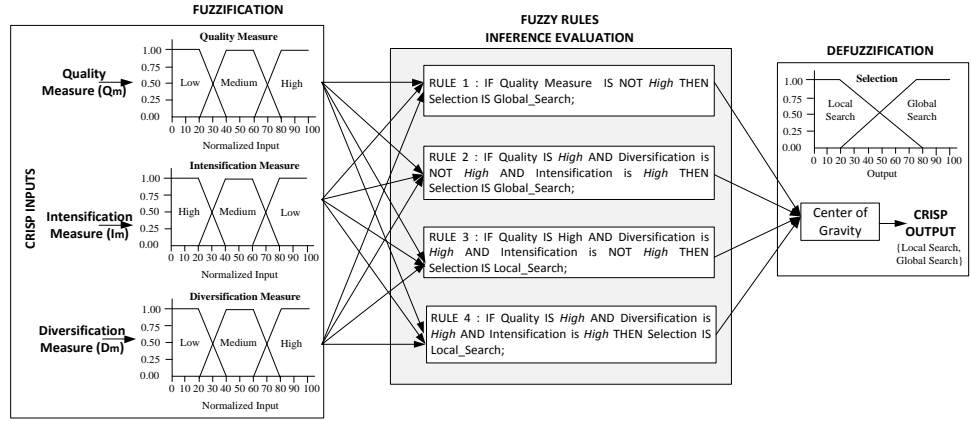

**Fig. 4.** Fuzzy inference system for ATLBO [20]

We opt to select three cases studies. Case study 1 relates to the printer manager software. Case study 2 relates to IOT controller whilst case study 3 deals with layer monitor. These case studies are actually students' class projects for undergraduate course. The complete class diagram description or all the case studies are shown in Fig. 5. Although the chosen case studies are rather small, it is enough to demonstrate the usefulness of our approach.

In all our experiments, we set the population size of 40, and the maximum fitness function evaluations of 5000. Our experimental platform comprises of a PC running Windows 10, CPU 2.9 GHz Intel Core i5, 16 GB 1867 MHz DDR3 RAM and a 512 MB of flash HDD. In all the experiments, we have executed ATLBO and TLBO 20 times to ensure statistical significance. The best MQ and its standard deviation along with the mean for each experiment is reported side-by-side. The best mean cell entries are marked as bold font.

## 5      Discussion

Table 1 summarizes the results. Referring to Table 1, we note that all TLBO variants can achieve the best MQ of 2.038, 2.011 and 1.500 respectively. ATLBO gives the best overall means for all three cases. For case study 3, ALTBO and FATLBO share the best performance. TLBO performs the poorest- with no entry achieving the best mean.

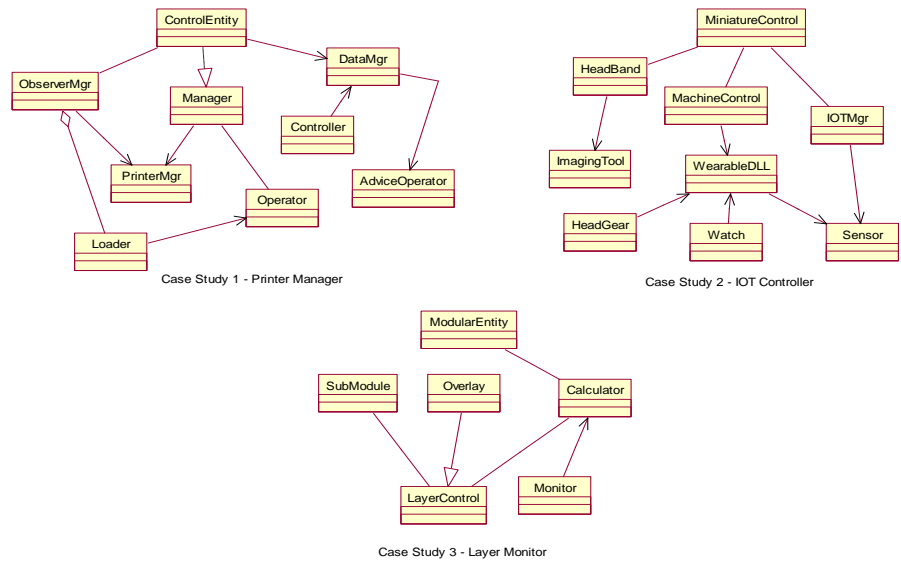

**Fig. 5.** Selected case studies

**Table 1.** Comparative MQ performance of ATLBO vs TLBO and FATLBO

| Case Studies | TLBO [21] | | ATLBO [20] | | FATLBO [19] | |
|---|---|---|---|---|---|---|
| | Best | Mean | Best | Mean | Best | Mean |
| Case Study 1: Printer Manager | 2.038±0.066 | 2.004 | 2.038±0.040 | **2.035** | 2.038±0.050 | 2.012 |
| Case Study 2: IOT Controller | 2.011±0.038 | 1.992 | 2.011±0.000 | **2.011** | 2.011±0.038 | 1.992 |
| Case Study 3: Layer Monitor | 1.500±0.175 | 1.2860 | 1.500±0.00 | **1.500** | 1.500±0.00 | **1.500** |

Concerning time performance, we note that all algorithms have similar performances. In fact, from our observation, there is no significant difference as far as execution time is concerned.

Referring to Fig. 6, we revisit the class diagram from Fig. 5. Based on the best output from ATLBO, we cluster all the classes based on the best MQ that are generated. Ideally, these clusters can be considered as packages or sub-modules. It must be stressed to have a MQ measure that finds balance between coupling and cohesion but not to completely remove them. For instance, one can have only 1 cluster or $n$ completely independent single module clusters to have zero coupling but such an approach does not aid in functionality-cluster-module traceability.

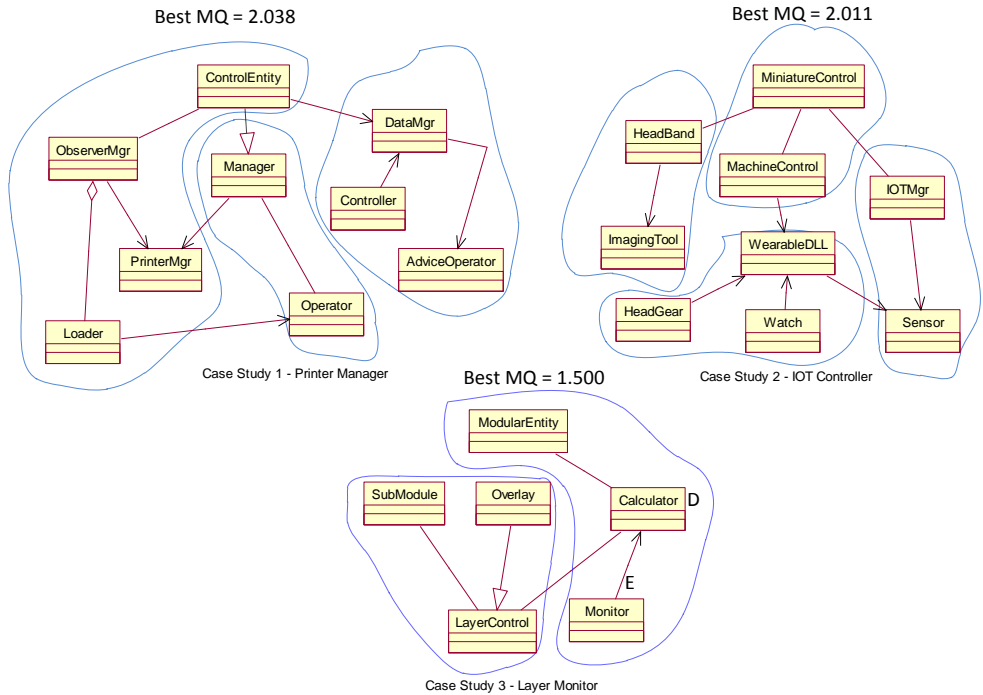

**Fig. 6.** Corresponding best clusters MQs for case studies 1 till 3

Finally, given both ATLBO and FATLBO adopt Mamdani fuzzy as their improvement to TLBO, it is necessary to elaborate their similarities and differences. FATBLO uses the success rate of both teacher phase and student phase as the fuzzy inputs. Essentially, FATLBO relies on penalize and reward scheme to ensure the execution of either teacher or learner phase (or both) via fuzzy probabilistic bar movement. Performing phase has higher probability for re-selection in the next iteration. ATLBO, on the other hand, relies on the quality measure, diversification measure and intensification measure to determine the best phase for selection. In this manner,

ATLBO appears more sensitive on the current needs at specific instance of the searching process as compared to FATLBO.

## 6    Conclusion

Summing up, this paper has demonstrated the effectiveness of ATLBO for software module clustering applications. ATLBO has outperformed FATLBO and the original TLBO in terms of generating the optimal MQ measure. As the scope of future work, we are exploring the adoption of ATLBO for other optimization problems particularly on wireless sensor network localization and software product line test suite generation.

## Acknowledgment

The work reported in this paper is funded by Fundamental Research Grant from Ministry of Higher Education Malaysia titled: A Reinforcement Learning Sine Cosine based Strategy for Combinatorial Test Suite Generation (grant no: RDU170103). We thank MOHE for the contribution and support. Fakhrud Din is the recipient of the Malaysian International Scholarship from the Ministry of Higher Education, Malaysia